\documentstyle[aps,eqsecnum]{revtex} 
\draft
\twocolumn
\begin{document}

\title{Squeezing enhancement by damping in a driven atom-cavity system}
\author{Hyunchul Nha$^{1,2}$, Young-Tak Chough$^2$, Sang Wook Kim$^2$, and Kyungwon An$^2$}
\address{$^1$School of Physics, Seoul National University, Seoul, Korea} 
\address{$^2$Center for Macroscopic Quantum-Field Lasers and Department of Physics, KAIST, Taejon, Korea} 
\date{\today}
\maketitle

\begin{abstract} 
In a driven atom-cavity coupled system in which the two-level atom is driven by a classical field, the cavity mode which should be in a coherent state in the absence of its reservoir,  can be squeezed by coupling to its  reservoir. The squeezing effect is enhanced as the damping rate of the cavity is increased to some extent.  
\end{abstract}
%\hspace{1.2cm}
%{\bf OCIS codes:} 270.6570, 270.5290, 270.2500

%\vspace{-0.3cm}

A quantum-mechanical harmonic oscillator undergoes continuous amplitude fluctuation even in its ground state. This fluctuation, also known as the vacuum fluctuation, arises from the nonvanishing commutation relation, $[a,a^{\dag}]=1$, where $a$ $(a^{\dag})$ is the annihilation (creation) operator of the harmonic oscillator. One can make, however, the fluctuation of one quadrature amplitude decrease below the vacuum-state (or the coherent state) level at the cost of that of the other quadrature.

This mechanism, known as squeezing \cite{Walls},  is most commonly generated by the inherent nonlinearity of the interaction involved, which is seen from the definition of the squeezing operator $S(r)=\exp[(r/2)\{(a^{\dag})^2-a^2\}]$, $r$ being the squeezing parameter. Squeezing mechanism may thus be interpreted as making pairs of correlated oscillator quanta. In most cases, the oscillator is inevitably damped by its coupling to the reservoir, and the squeezing is degraded since the coupling introduces the reservoir fluctuations to the oscillator. For instance, the known schemes for generating the squeezed light employ the nonlinear processes \cite{phonon} such as parametric amplification \cite{Zubairy} or four-wave mixing \cite{Slusher}. Usually, the interacting medium is put inside an optical cavity in order to build up the field by increasing the interaction time \cite{Zubairy}.  But the cavity damping destroys the quantum correlation of the field via loss of the correlated photon pairs.

In the conventional squeezing schemes, therefore, damping has been considered to play a negative role. In this work, by contrast, we show that squeezing is manifested by damping in a counter-intuitive way, when the cavity is coupled to its reservoir under {\it indirect pumping}. The coupling strengths to each mode of the reservoir oscillators are related to the damping rate, $\kappa$, of the cavity mode by the fluctuation-dissipation theorem \cite{Huang}.  We will show that the squeezing effect is enhanced as $\kappa$ is increased to some degree.

In our model, the target harmonic oscillator, i.e., the cavity mode, is coupled to a two-level atom  driven by a resonant classical field, instead of being pumped directly. Although the system has the inherent nonlinear property due to the saturability of the two-level atom, the nonlinearity producing the squeezing effect does not reveal itself without the cavity damping. In addition, the unconventional features of the squeezing enhancement by damping can be seen from its characteristic condition $2\Omega\kappa/g^2\sim1$, where $g$ is the atom-cavity coupling constant and $\Omega$,  the Rabi frequency of the driving field. This condition shows that the Rabi frequency should be {\it inversely} proportional to the cavity damping rate for the optimal squeezing. This relation is in a striking contrast to those of the conventional schemes, where the pumping intensity should be proportional to the damping rate.

We employ the standard master equation in order to consider appropriately the cavity damping as well as the atomic damping in calculating numerically the dynamics of the system. 
To begin with, our system is described by the interaction Hamiltonian  
\begin{eqnarray}
H_I=i\hbar g (a^{\dag}\sigma_{-}-a\sigma_{+})+i\hbar\Omega(\sigma_+-\sigma_-) 
\end{eqnarray}
where $\sigma_{\pm}$ and $\sigma_z$ are the atomic pseudospin operators and ${a}^{\dag}(a)$, the creation(annihilation) operator of the cavity mode. We find the steady-state behavior of the system in the presence of the dampings from the master equation which in the interaction picture is given by
\begin{eqnarray}
\dot{\rho_I}&=&\frac{1}{i\hbar}[H_I,\rho_I]+\kappa(2a\rho_I a^{\dag}-a^{\dag}a\rho_I-\rho_I a^{\dag}a)\nonumber\\
&+&(\gamma/2)(2\sigma_-\rho_I \sigma_+-\sigma_+\sigma_-\rho_I-\rho_I \sigma_+\sigma_-),
\end{eqnarray}
where $\gamma$ is the decay rate of the atom into free-space vacuum-field modes. In the present form, the cavity damping may seem to play the usual phase-insensitive role, only to degrade the squeezing which is the phase-sensitive mechanism. However, if we transform the above equation using $\tilde{\rho} =D^+(\alpha)\rho_ID(\alpha)$, where $D(\alpha)$ is the displacement operator for the cavity-mode with $\alpha=\Omega/g$, into
\begin{eqnarray}
\dot{\tilde{\rho}}&=&[g(a^{\dag}\sigma_{-}-a\sigma_{+}),\tilde{\rho}]+\Omega\frac{\kappa}{g}
[a-a^{\dag},\tilde{\rho}]\nonumber\\
&+&\kappa(2a\tilde{\rho} a^{\dag}-a^{\dag}a\tilde{\rho}-\tilde{\rho} 
a^{\dag}a)\nonumber\\
&+&(\gamma/2)(2\sigma_-\tilde{\rho} \sigma_+-\sigma_+\sigma_-\tilde{\rho}-\tilde{\rho} \sigma_+\sigma_-),
\end{eqnarray}
completely new aspect of the cavity damping starts to emerge \cite{nha}.  In Eq.\,(3), we see that the cavity damping appears in two different  places. In the first place, the cavity damping assumes an anomalous role of causing the squeezing while in the second it plays the usual  role of damping, as clearly demonstrated below.

Note first that in our displaced frame  we have  an effective Hamiltonian 
\begin{eqnarray}
H_{\rm eff}=i\hbar g(a^{\dag}\sigma_{-}-a\sigma_{+})+i\hbar\Omega\frac{\kappa}{g}(a-a^{\dag})
\end{eqnarray} 
which is the very interaction Hamiltonian of a system where the cavity mode is directly driven by a resonant driving-field of strength ${\cal E}_{\rm eff}=\Omega\kappa/g$. Note that ${\cal E}_{\rm eff}$ is proportional to the damping rate $\kappa$. This Hamiltonian explains all the features involved in the ``squeezing enhancement by damping.'' The eigen-energies and the corresponding eigen-states for the Hamiltonian of the above type were calculated by Alsing {\it et al.} \cite{Alsing1}, which we quote below. The ground state of $H_{\rm eff}$ is given by $S(r)|0\rangle|A^-\rangle$, where $S(r)=\exp\{(r/2)[(a^{\dag})^2-a^2]\}$ is the squeezing operator for the cavity mode, and $|A^-\rangle=1/\sqrt{2}(\epsilon^+|-\rangle-\epsilon^-|+\rangle)$ with $\epsilon^{\pm}=(1\pm e^{2r})^{1/2}$. The state $|0\rangle$ refers to the vacuum state of the cavity mode and $|\pm\rangle$ to the atomic excited/ground state. In our case, the squeezing parameter $r$ is given by
\begin{eqnarray}
e^{2r}=\sqrt{1-(2\Omega\kappa/g^2)^2}.
\end{eqnarray}
As we return to the interaction picture, we find that the state of cavity mode is a squeezed coherent state $D(\alpha)S(r)|0\rangle$, which is also the minimum-uncertainty squeezed state. 

Now, for comparison, let us consider the case of loss-less cavity, i.e., $\kappa=0$. Then, with the squeezing parameter $r=0$, the ground state of the coupled system becomes $\rho_I^{ss}=|\alpha\rangle\langle\alpha|_a|-\rangle\langle-|_b$, which is also the steady state of the system in the presence of nonzero $\gamma$ \cite{Alsing2}. In this case, the cavity field is in a coherent state of amplitude $\alpha$ and the atom in its ground state. Thus we see that only when the cavity damping $\kappa$ is nonzero, the cavity mode can be squeezed and the atom gets excited.

We can see from Eq.\,(5) that the squeezing effect is induced more strongly as $\kappa$ is increased.  If the state of the cavity mode is decomposed in the number state basis, the number distribution is sub-Poissonian because the effective driving field ${\cal E}_{\rm eff}$ induces the quadrature squeezing in $X_1=(a+a^{\dag})/2$ with $r<0$ in Eq.\,(5) \cite{Scully}. 
Thus, the Mandel-$Q$ factor defined as $Q=\Delta n^2/\langle n\rangle-1$ decreases as $\kappa$ increases for a given driving-intensity $\Omega$, as shown in Fig.\,1 (a). In addition, the fluctuation in quadrature $X_1$ of oscillator $a$ decreases below the vacuum level given by $4\Delta X_1^2=1$. [see Fig.\,1 (b).] The strongest sub-Poissonian distribution and quadrature squeezing are obtained roughly for
\begin{eqnarray}
2\Omega\kappa/g^2={\rm const.}\sim1,
\end{eqnarray}
i.e., when the squeezing parameter $r$ becomes negative infinite ($e^{2r}\approx0$) \cite{nha1}.

The mean excitation number for the cavity field decreases monotonically as $\kappa$ increases, as checked 
in Fig.\,2 (a).  It is noteworthy, however, that the mean excitation number is not so critically changed from the 
value $|\Omega/g|^2$ (of $\kappa=0$ case) as a function of $\kappa$, until the condition $2\Omega\kappa/g^2\sim1$ 
is reached, although the variance $\Delta n$ of the number decreases significantly (not shown). In fact, this is 
why the Mandel-$Q$ factor is minimal when $2\Omega\kappa/g^2\sim1$. Unlike other squeezing schemes based on 
phase-selective amplification, where squeezing increases the mean excitation number \cite{carmichael}, in this case, 
the excitation number cannot really increase due to the usual damping effect, as seen in the second line in Eq.\,(3), and it 
decreases adiabatically as the cavity mode gets more strongly squeezed.

Although the damping decreases the fluctuation of one quadrature $X_1$ only to the extent $2\Omega\kappa/g^2\sim1$, 
it always increases the phase-fluctuation of the oscillator because the cavity damping always causes 
 decoherence in the cavity mode. When a phase operator is introduced as 
$\hat{\cos}\phi=[(a^\dag a+1)^{-\frac{1}{2}}a+a^\dag(a^\dag a+1)^{-\frac{1}{2}})/2$ \cite{Susskind}, the fluctuation 
$\Delta\hat{\cos}\phi$ increases monotonically in proportion to the damping rate, as seen in Fig.\,2 (a). In the 
extreme case of $\kappa/g\gg1$, the cavity mode approaches the vacuum state $|0\rangle$, so the phase fluctuation 
$\Delta\hat{\cos}\phi\rightarrow0.5$.

Now, turning our attention to the cavity-output field \cite{Yurke,Car}, we can numerically calculate the spectrum of 
squeezing, according to 
$S(\omega,\theta)=16\Gamma_a\int_0^\infty d\tau\cos\omega\tau\langle:\Delta X_{\theta}(0)\Delta X_{\theta}(\tau):
\rangle$ \cite{collet}. The best squeezing is obtained near the frequency 
$\omega-\omega_0=\pm g[1-(2\Omega\kappa/g^2)^2]^{3/4}$ (see Fig.\,3), which reflects the eigenvalues of the 
Hamiltonian $H_{\rm eff}$ that we derived, given by $E_n^{\pm}=\pm\hbar\sqrt{n}g[1-(2\Omega\kappa/g^2)^2]^{3/4}$, 
$(n=0, 1, \cdots)$. 

In conclusion, we investigated a squeezing effect manifested by damping in the driven atom-cavity system. It has 
been shown that the squeezing effect is enhanced as the damping rate of the cavity is increased to some extent. 
Furthermore, the pumping field amplitude is required to be inversely proportional to the damping rate for the 
optimal squeezing. Our findings are well confimed in the numerical simulations based on the master 
equation in its original form, Eq.\,(2). This novel effect of damping revealed here is highly counter-intuitive 
and somewhat reminiscent of the ``noise suppression by noise'' recently reported in the classical statistical 
physics \cite{freezing-by-heating}. 

We thank H. J. Carmichael for helpful discussions. This work is supported by Creative Research Initiatives of the Korean Ministry of Science and Technology.

%\vspace{-0.5cm}

\begin{figure}
%\vspace{2.8in}
%\special{pict=fig1.pict2 scale=0.24}
\caption{(a) The Mandel-$Q$ value, and (b) $4\Delta X_1^2$ as a function of 
$2\Gamma_a/g$ and $\Omega/g$ for $\Gamma_b/g=0.01$. From the contour plots below, it is seen that the 
squeezing effect is enhanced as $\Gamma_a$ is increased, for a given driving-intensity, to 
a certain value roughly given as $2\Omega\Gamma_a/g^2=c\sim1$ (dotted line).}
\end{figure}

\begin{figure}
%\vspace{1.5in}
%\special{pict=fig2.pict scale=0.115}
\caption{(a) Mean excitation number (divided by 4), the Mandel-$Q$ value, and the phase fluctuation $\Delta \hat{\cos}\phi$. (b) $\Delta X_1$, and 
$\Delta X_2$ as a function of $2\Gamma_a/g$ for 
$\Omega/g=2$,$\Gamma_b/g=0.02$. When the damping is very small, the ideal squeezed state 
$(4\Delta X_1\Delta X_2=1)$ is obtained.} 
\end{figure}

\begin{figure}
%\vspace{2.7in}
%\special{pict=fig3.pict scale=0.175}
\caption{(a) Spectrum of squeezing for $X_1$ and $X_2$ for 
$\Omega/g=0.53$, $2\Gamma_a/g=1.33$, and $2\Gamma_b/g=0.02$. }
\end{figure}


\begin{references}
\bibitem{nha1}The larger atomic damping always causes the degradation of the squeezing, contrary to the cavity damping, which was confirmed independently in the numerical simulations.
\end{references}
\end{document}